\documentclass[twocolumn]{aastex631}
\usepackage{amsmath}

\shorttitle{Galactic Exhaust Vent}
\shortauthors{Mackey et al.}

\graphicspath{{./}{Figures/}}

\begin{document}

\title{X-rays from a Central “Exhaust Vent” of 
the Galactic Center Chimney}

\submitjournal{ApJ}
\received{2023 September 29}

\correspondingauthor{Scott C. Mackey}
\email{scmackey@uchicago.edu}

\author[0000-0002-1414-7236]{Scott C. Mackey}
\affiliation{Department of Physics \& Kavli Institute for Cosmological Physics, University of Chicago, Chicago, IL 60637, USA}
\affiliation{Department of Physics and Astronomy,
University of California, Los Angeles, CA 90095, USA}

\author[0000-0002-6753-2066]{Mark R. Morris}
\affiliation{Department of Physics and Astronomy,
University of California, Los Angeles, CA 90095, USA}

\author{Gabriele Ponti}
\affiliation{INAF Osservatorio Astronomico di Brera, Via Bianchi 46, I-23807 Merate (LC), Italy}
\affiliation{Max-Planck-Institut für extraterrestrische Physik, D-85748 Garching, Germany}

\author{Konstantina Anastasopoulou}
\affiliation{INAF Osservatorio Astronomico di Palermo, Piazza del Parlamento 1, I-90134 Palermo, Italy}

\author{Samaresh Mondal}
\affiliation{INAF Osservatorio Astronomico di Brera, Via Bianchi 46, I-23807 Merate (LC), Italy}



\begin{abstract}

 Using deep archival observations from the Chandra X-ray Observatory, we present an analysis of linear X-ray-emitting features located within the southern portion of the Galactic center chimney, and oriented orthogonal to the Galactic plane, centered at coordinates $l = 0.08^{\circ}, ~b = -1.42^{\circ}$. The surface brightness and hardness ratio patterns are suggestive of a cylindrical morphology which may have been produced by a plasma outflow channel extending from the Galactic center. Our fits of the feature's spectra favor a complex two-component model consisting of thermal and recombining plasma components, possibly a sign of shock compression or heating of the interstellar medium by outflowing material. Assuming a recombining plasma scenario, we further estimate the cooling timescale of this plasma to be on the order of a few hundred to thousands of years, leading us to speculate that a sequence of accretion events onto the Galactic Black Hole may be a plausible quasi-continuous energy source to sustain the observed morphology.
 

\end{abstract}




\section{Introduction} \label{sec:intro}

Recent large-scale surveys of the Galactic center have revealed that the X-ray and radio emission on scales of several hundred parsecs assumes a bipolar morphology centered on the Galactic center, with the well-defined emitting lobes oriented perpendicular to the Galactic plane \citep{Ponti+19, Ponti+21, Heywood+19}.  These findings had been presaged by a considerable amount of earlier work over several decades on the ``Galactic Center Lobe" seen at northern galactic latitudes at radio wavelengths \citep{Sofue+Handa84, Sofue85, Law2010} and by X-ray observations of features extending to latitudes well out of the Galactic plane \citep{Nakashima+13, Ponti_soft15}.  The interpretation of the morphology by \citet{Ponti+21} is that the lobes are bubbles of hot plasma surrounded by a radio-emitting shell of denser ionized gas. \citet{Ponti+19} hypothesized that these bubbles are essentially ``chimneys" along which hot plasma is moving vertically out of the Galactic center where the plasma was produced, and into the region of the galaxy-scale Fermi gamma-ray bubbles \citep{Su+10, Yang+18,Yang+22} and the even larger X-ray bubbles detected by the eRosita X-ray observatory \citep{Predehl+20}.  The chimneys could thereby be the channels by which sources in the Galactic center have provided the energy and particles to feed the Fermi and eRosita bubbles.  However, evidence for actual motion of plasma along the chimneys has not yet been reported, and it remains unclear whether the galaxy-scale bubbles are fed primarily by relativistic particles, i.e., cosmic rays, diffusing up from the region of massive star formation near the Galactic plane, or by a hot, outflowing plasma driven either by extreme episodes of accretion onto the Galactic Black Hole or by the collective deposition of energy by many supernovae over time \citep{Zubovas+11, Zubovas+12, Crocker+Aharonian11, Crocker+15, ZhangMF+21}.  

Whatever the mechanisms for producing the large-scale bubbles might be, the energy generated in the central few hundred parsecs of the Galaxy by accretion onto the central supermassive black hole and by supernovae and stellar winds associated with recent star formation has clearly created a Galactic wind \citep{Crocker+10, Carretti+13, McClure-Griffiths+13, Fox+15, Ashley+20, Lockman+20}.  The driving mechanism for the wind from the Galactic center is likely tied to the mechanisms producing the Fermi and eRosita Bubbles. The wind launches material into the active circumgalactic medium, the arena of the Galactic fountain \citep{Tumlinson+17}.   

\begin{figure*}[ht!]
    \centering
    \includegraphics[width=400pt]{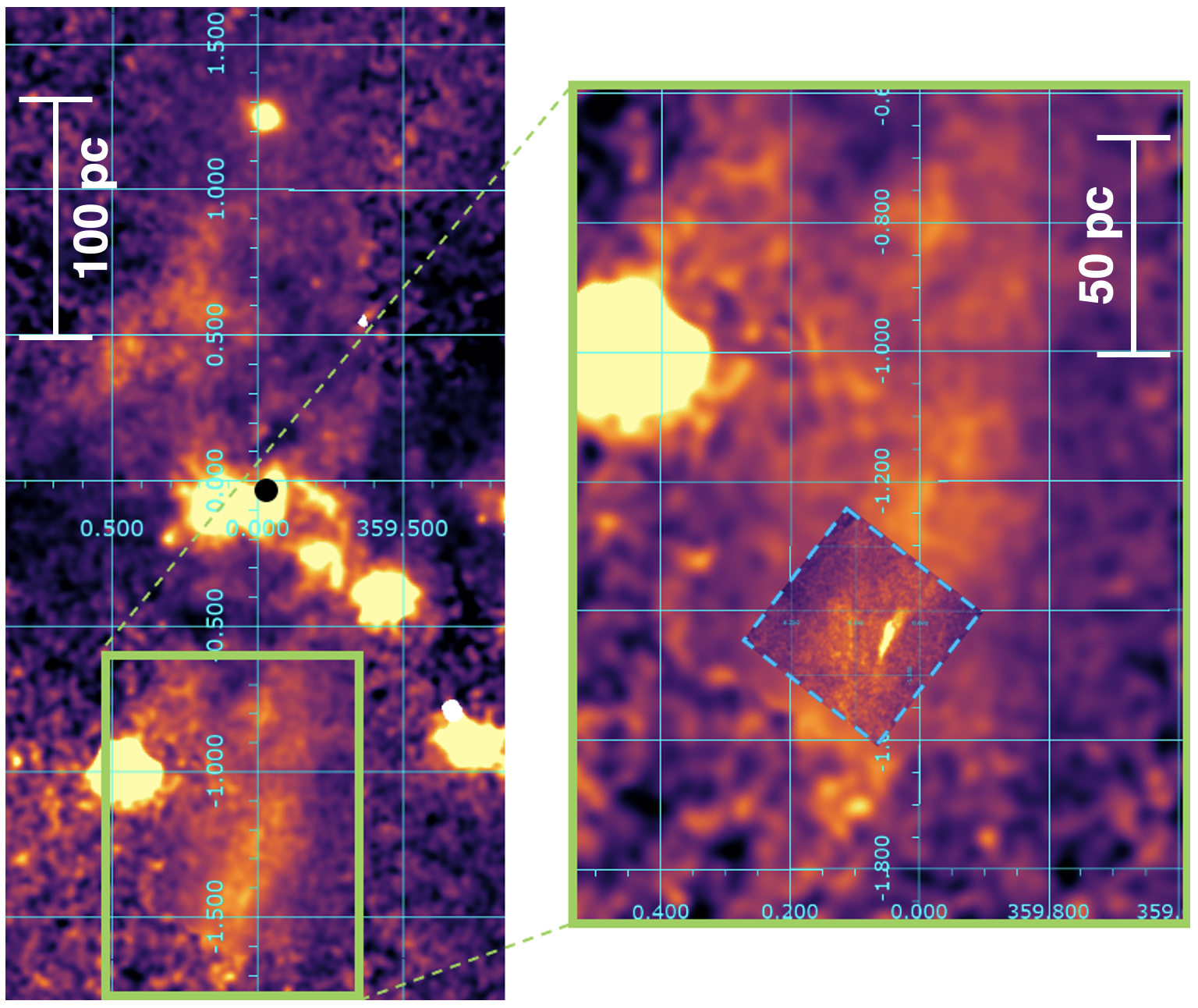}
    \caption{1.5-2.6 keV Chandra map of the Galactic center in Galactic coordinates, showing the chimneys. A 0.5-2.0 keV image of the Inner Bulge Deep Field is overlaid in the right panel and highlighted with blue dashed lines. The black dot in the left panel denotes the location of Sgr A*.}
    \label{fig:chimneys}
\end{figure*}

This paper presents an investigation of diffuse, extended X-ray emission from a single archival Chandra field that contains structures that appear morphologically to constitute the central vent in the southern Galactic center chimney, as shown in Figure \ref{fig:chimneys}. 
Observations and characteristics of this particularly deep field are described in section \ref{sec:methods}, while the resulting morphology of the emitting structures and a spectroscopic investigation of those structures are presented in section \ref{sec:results}. 
Finally, our interpretation of the observational results is given and discussed in section \ref{sec:disc}.  \\

\section{Observations and Data Preparation} \label{sec:methods}

\begin{figure*}[ht!]
    \centering
    \includegraphics[width=400pt]{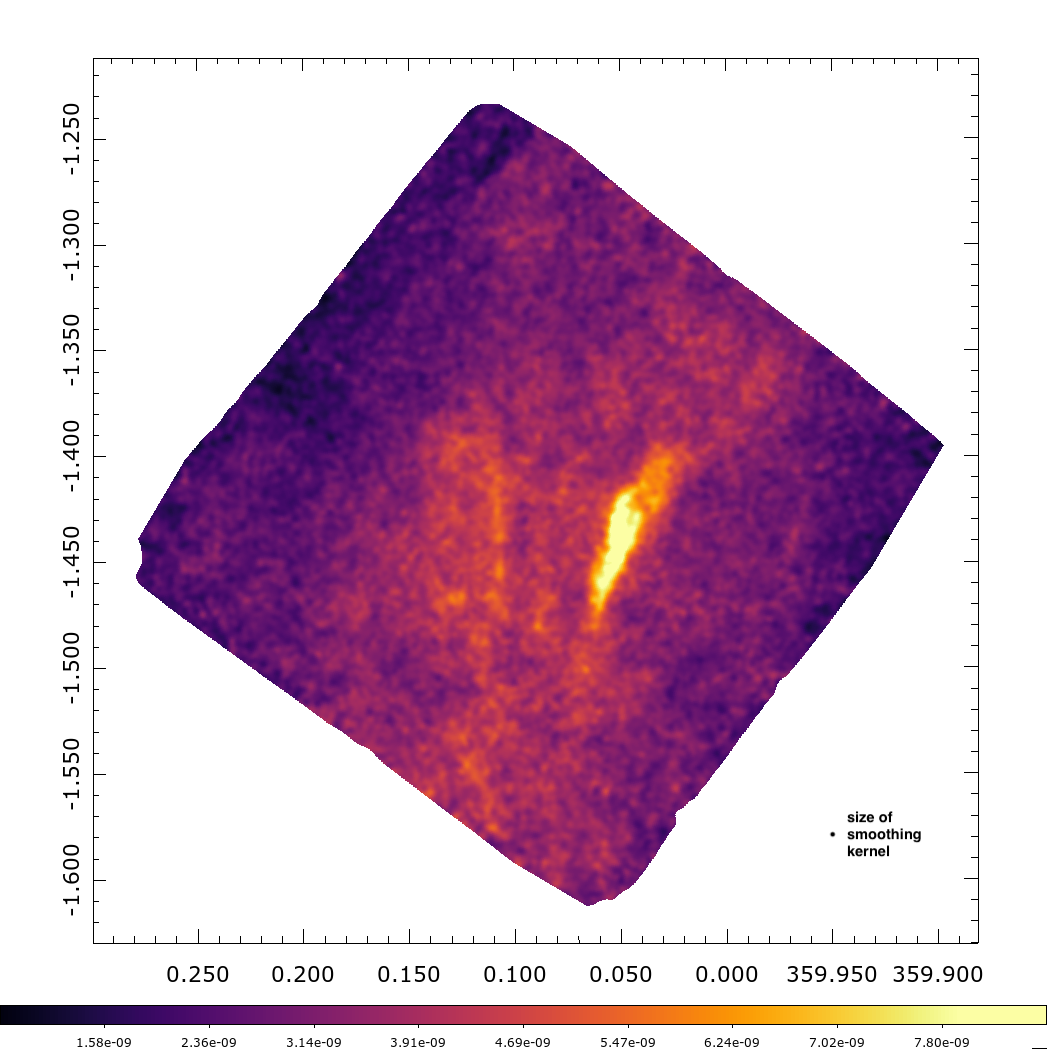}
    \caption{0.5-2 keV Chandra map of the IBDF. Galactic coordinates are shown in degrees.}
    \label{fig:xray}
\end{figure*}

For this study, we focus on a $\sim 17 \times 17$ arcmin field located at $l = 0.08^{\circ}, ~b = -1.42^{\circ}$, which we refer to as the Inner Bulge Deep Field (IBDF) (Figure \ref{fig:xray}). 
A total of 13 archival Chandra observations cover the field with a combined integration time of nearly 1 Ms (Table \ref{tab:obs}).
The majority of the observations were conducted by \cite{Revnivtsev+09} in 2008 to characterize the discrete point sources that are part of the Galactic ridge X-ray emission (GRXE), a continuous unresolved hard X-ray emission feature extending along the Galactic plane to longitudes of  $\pm\sim40^{\circ}$ \citep{Worrall+82, Warwick+85}. 
In addition, three of the observations were made in 2005 as part of the ChaMPlane Galactic bulge survey \citep{Vandenberg+09}.
All data were taken with the imaging array of the Advanced CCD Imaging Spectrometer (ACIS-I) onboard the Chandra X-ray Observatory. 
Data were downloaded from the Chandra Data Archive and reprocessed using the standard procedure for the Chandra Interactive Analysis of Observations (CIAO)\footnote{\url{https://cxc.harvard.edu/ciao/}} software package (version 4.13, CALDB 4.9.5), starting with the \texttt{chandra\_repro} script.

Interpreting diffuse X-ray emission requires a thorough treatment of the non-X-ray instrumental background.
We used the ACIS-I stowed background data set and re-scaled it by assuming that nearly all of the 9-12 keV flux is due to particle background, using a method based on the one prescribed by \cite{Hickox+06}.
Our method differs slightly from the background subtraction techniques previously used in similar analyses and is detailed in Appendix \ref{appendix:a}.
After subtracting this background from each observation, we combined and exposure-corrected all observations to produce a mosaic image.
Since the roll angles vary among the various observations, the images do not always align well, which results in a few regions where there is little or no overlap with the bulk of the observations and therefore drastically lower exposure times. We addressed this by simply masking such regions out of the image.
To remove point sources, we identified them using CIAO's \texttt{wavdetect} routine with a sensitivity threshold (i.e. tolerance of falsely identifying a pixel as belonging to a source) of $1\times10^{-6}$ and then excised and interpolated across the source regions using \texttt{dmfilth}.
Finally, the image was smoothed using a constant Gaussian kernel with a radius of 6 pixels (3 arcsec).

Spectra were produced using \texttt{specextract} and then fit using the XSPEC \footnote{\url{https://heasarc.gsfc.nasa.gov/xanadu/xspec/}} software package \citep{Arnaud96}. 
Because the X-ray features studied here are dominated by emission below 2 keV and the diffuse cosmic X-ray background is best accounted for up to about 4 keV \citep{Hickox+06,Hickox+07}, we limit all of our spectral analyses to energies below 3 keV. 
Instrumental background was subtracted from the spectra before fitting, and spectra were rescaled based on area.
We find that the fits show little dependence on elemental abundances, so we fix this to solar abundance.
The abundance table provided by \cite{Anders+89} was assumed.
We represent a thermal plasma with the \texttt{apec} (Astrophysical Plasma  Emission Code; hereafter `APEC') model and a non-equilibrium recombining plasma with the \texttt{rnei} (hereafter `RP') model.
Both models are default models within the XSPEC software package.

\begin{deluxetable*}{ccc}
\tablenum{1}
\tablecaption{Observations\label{tab:obs}}
\tablewidth{0pt}
\tablehead{
\colhead{ObsID} & \colhead{Date} & \colhead{Exposure} \\
\colhead{} & \colhead{(yyyy-mm-dd)} & \colhead{(ks)}
}
\startdata
5934 & 2005-08-22 & 40.49 \\
6362 & 2005-08-19 & 37.70 \\
6365 & 2005-10-25 & 20.69 \\
9500 & 2008-07-20 & 162.56 \\
9501 & 2008-07-23 & 131.01 \\
9502 & 2008-07-17 & 164.12 \\
9503 & 2008-07-28 & 102.31 \\
9504 & 2008-08-02 & 125.42 \\
9505 & 2008-05-07 & 10.73 \\
9854 & 2008-07-27 & 22.77 \\
9855 & 2008-05-08 & 55.94 \\
9892 & 2008-07-31 & 65.79 \\
9893 & 2008-08-01 & 42.16 \\
\enddata
\end{deluxetable*}

\section{Results} 
\label{sec:results}
\subsection{Morphology}
\begin{figure*}[ht!]
    \centering
    \includegraphics[width=300pt]{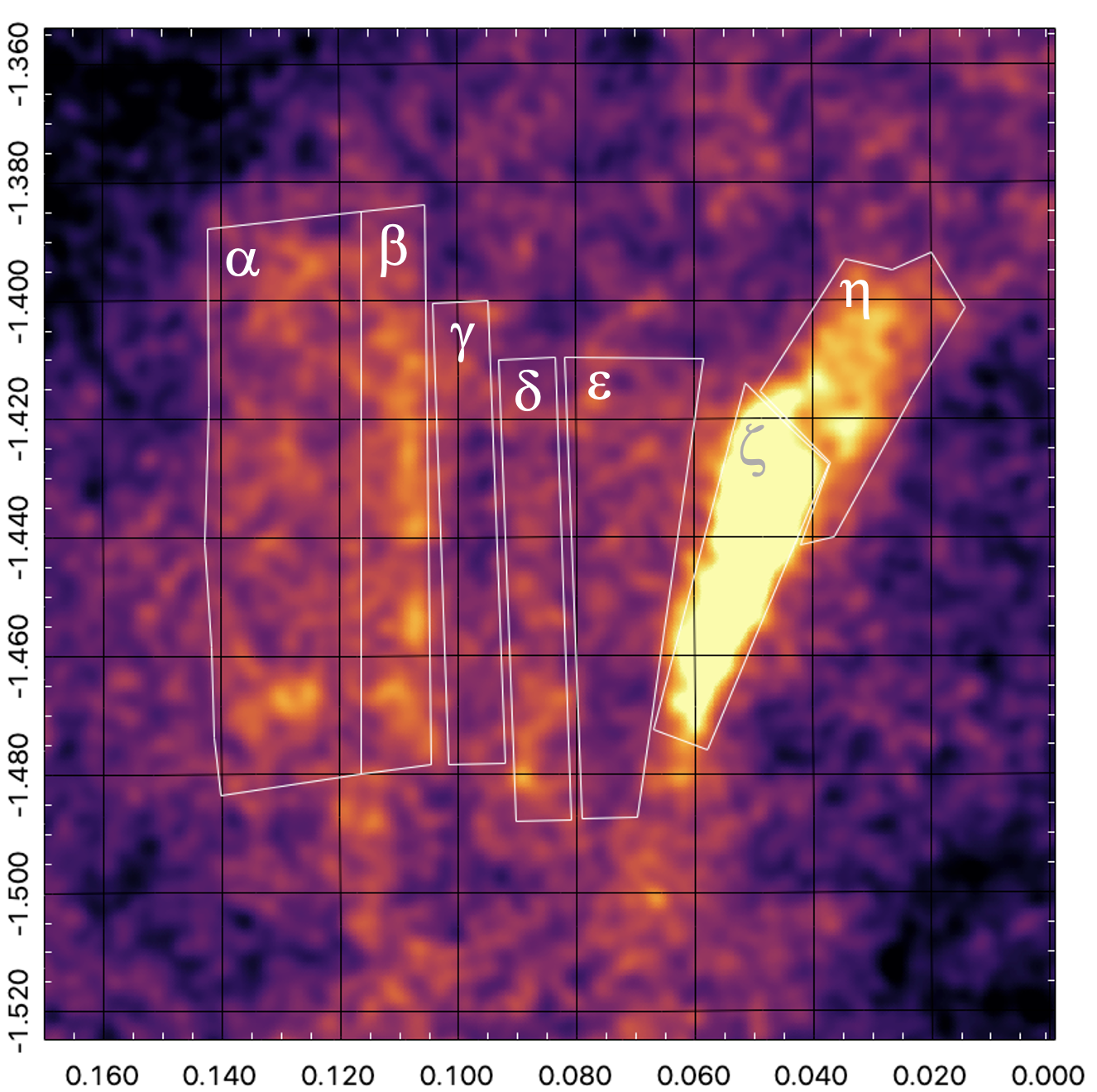}
    \caption{IBDF regions used for analysis. Galactic coordinates shown.}
    \label{fig:reg}
\end{figure*}

The features of interest within the IBDF are bright, linear ridges of X-ray emission oriented roughly perpendicular to the Galactic plane. 
They are characterized by relatively soft X-rays and do not appear outside of the 0.5-2.0 keV band.
We visually define seven regions across the IBDF field denoted by Greek letters $\alpha$ through $\eta$ going from positive to negative Galactic longitudes (Figure \ref{fig:reg}).

Feature $\beta$ is a narrow, bright ridge that runs parallel to lines of constant Galactic longitude and points toward the Sagittarius A complex at the Galactic center. 
Features $\gamma$, $\delta$, and $\epsilon$ are parallel to $\beta$, and while they are noticeably dimmer than $\beta$, they are regularly spaced in longitude and alternate in surface brightness, proceeding from $\beta$ to $\zeta$, so that feature $\delta$, situated at the center of the seven regions, appears brighter than the regions on either side of it. 
Regions $\zeta$ and $\eta$ combine into a distinct feature of notable complexity, as can best be seen in Figure \ref{fig:reg}. 
Its orientation is tilted with respect to that of the other parallel features by a small angle ($\sim 15^{\circ}$) and it features a brighter component ($\zeta$) on its eastern side bordering $\epsilon$ and a dimmer component ($\eta$) on its western side. 

\begin{figure*}[ht!]
    \centering
    \includegraphics[width=500pt]{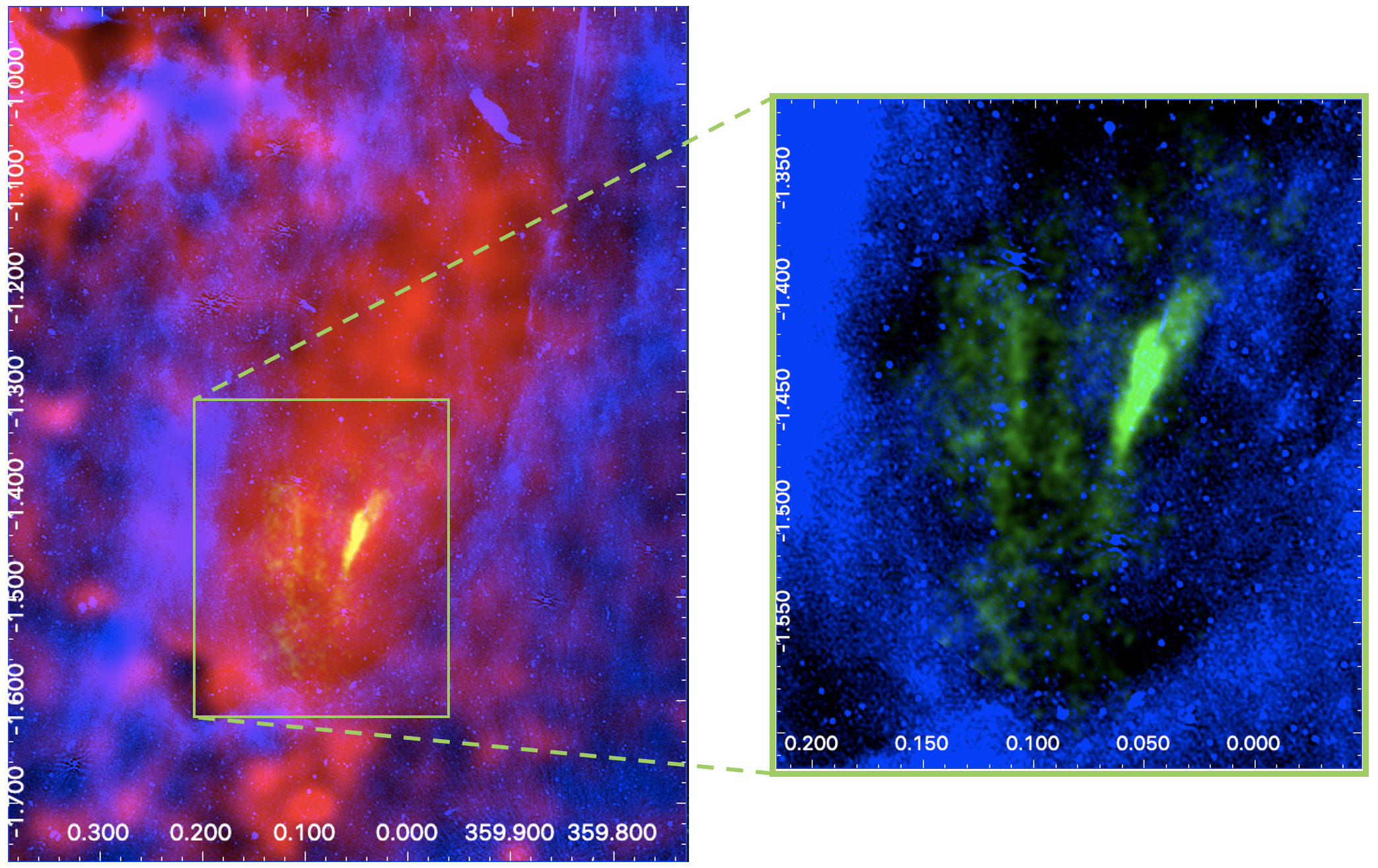}
    \caption{Multi-wavelength image of the southern radio lobe and chimney.  The IBDF X-ray features are shown in green (creating yellow in the left panel when mixed with red; 0.5-2.0 keV), the chimney diffuse X-ray emission is shown in red (1.5-2.6 keV), and the 1.284 GHz MeerKAT radio map is shown in blue. Emission from the chimney is omitted from the right panel to show the distribution of radio emission within the IBDF X-ray features.} 
    \label{fig:radio}

\end{figure*}

The IBDF X-ray features are located within the southern Galactic chimney, which itself is nestled within the southern radio lobe \citep{Ponti+19, Ponti+21, Heywood+19}, as shown in Figure \ref{fig:radio}.  As was noted by \citet{Ponti+21}, the radio emission arises from the periphery of the southern X-ray chimney, which evokes a structure in which the X-rays arise from a confined hot plasma, while the radio emission arises in a thick, higher-density surrounding shell where the hot plasma interacts with the ambient interstellar medium and the predominantly vertical magnetic field of the Galactic center.

\begin{figure*}[h]
    \centering
    \includegraphics[width=450pt]{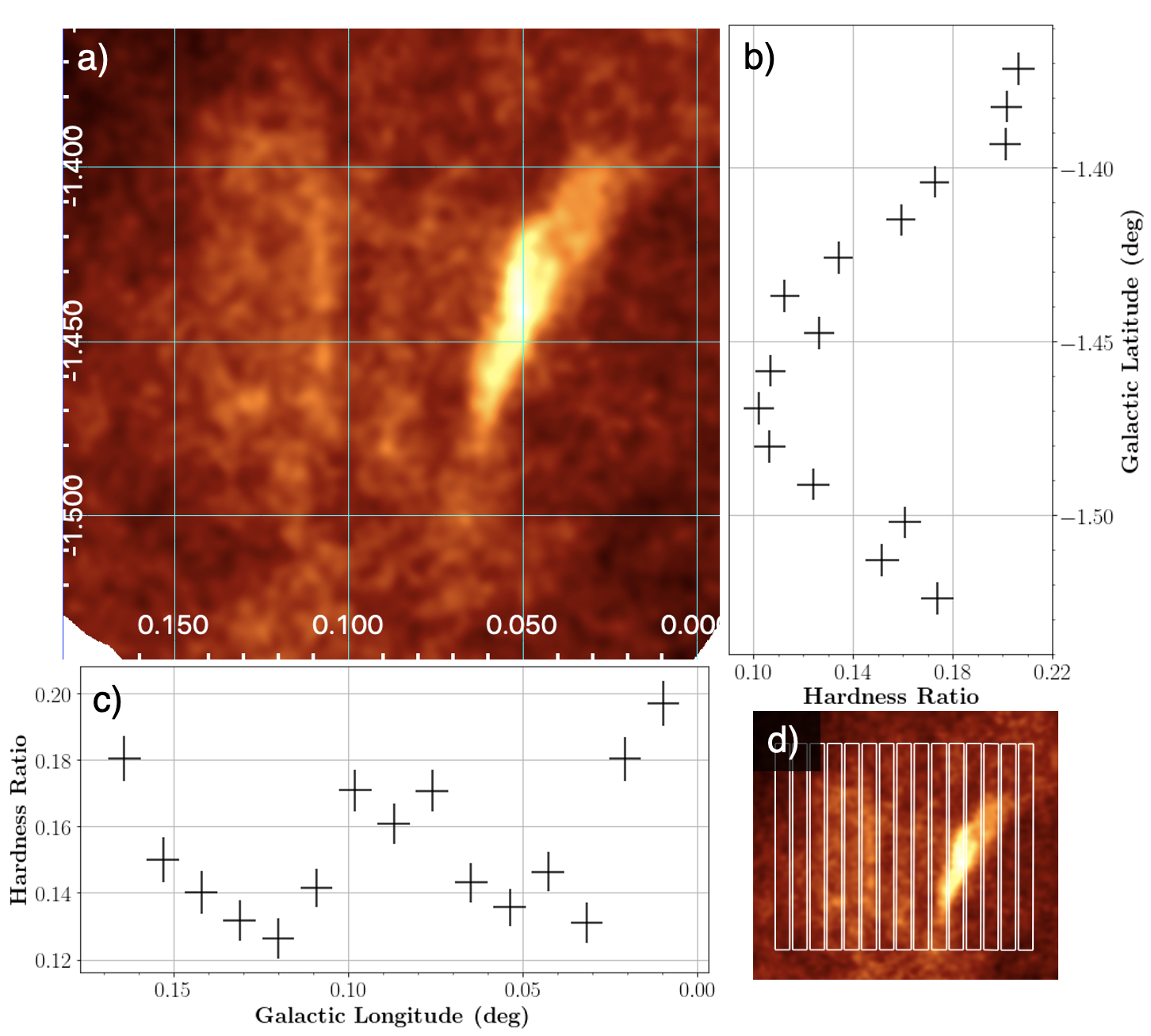}
    \caption{Hardness ratio of 1.2-2.0 keV to 0.5-1.2 keV emission.}
    \label{fig:hr1}
\end{figure*}

\subsection{Hardness Ratios}
The hardness ratio comparing the 1.2-2.0 keV band to the 0.5-1.2 keV band across the field is shown in Figure \ref{fig:hr1}.
We define the hardness ratio as
\begin{equation}
    \mathrm{Hardness \ Ratio} = \frac{C_{\mathrm{high}} - C_{\mathrm{low}}}{C_{\mathrm{high}} + C_{\mathrm{low}}}
\end{equation}
where $C_{\mathrm{high}}$ and $C_{\mathrm{low}}$ are the counts in the higher and lower energy bands, respectively, and a higher hardness ratio indicates harder emission.
The error bars for latitude and longitude represent the uniform 0.01 degree-wide segments over which the counts were summed, while the hardness ratio error comes from the statistical error of the observed counts in these segments.
Progressing across the field in Galactic longitude, the brighter linear features $\beta$, $\delta$, and $\zeta$ each roughly correspond to lower hardness ratios, while the overall minimum hardness ratio occurs in region $\alpha$.
In the progression with Galactic latitudes, the hardness ratio shows a broad minimum at the bright centers of the IBDF, with a small peak at $b = -1.448 \pm 0.005$ degrees and a broad minimum at $b = -1.469 \pm 0.005$ degrees.
Panel C in Figure \ref{fig:hr1} demonstrates differences in X-ray hardness within the $\zeta$/$\eta$ feature, with the softer eastern edge corresponding to the bright ridge seen in Figure \ref{fig:xray}, and relatively hard emission trailing off to the west.

\subsection{Spectra}
\label{sec:spectra}
\begin{figure*}[ht!]
    \centering
    \includegraphics[width=450pt]{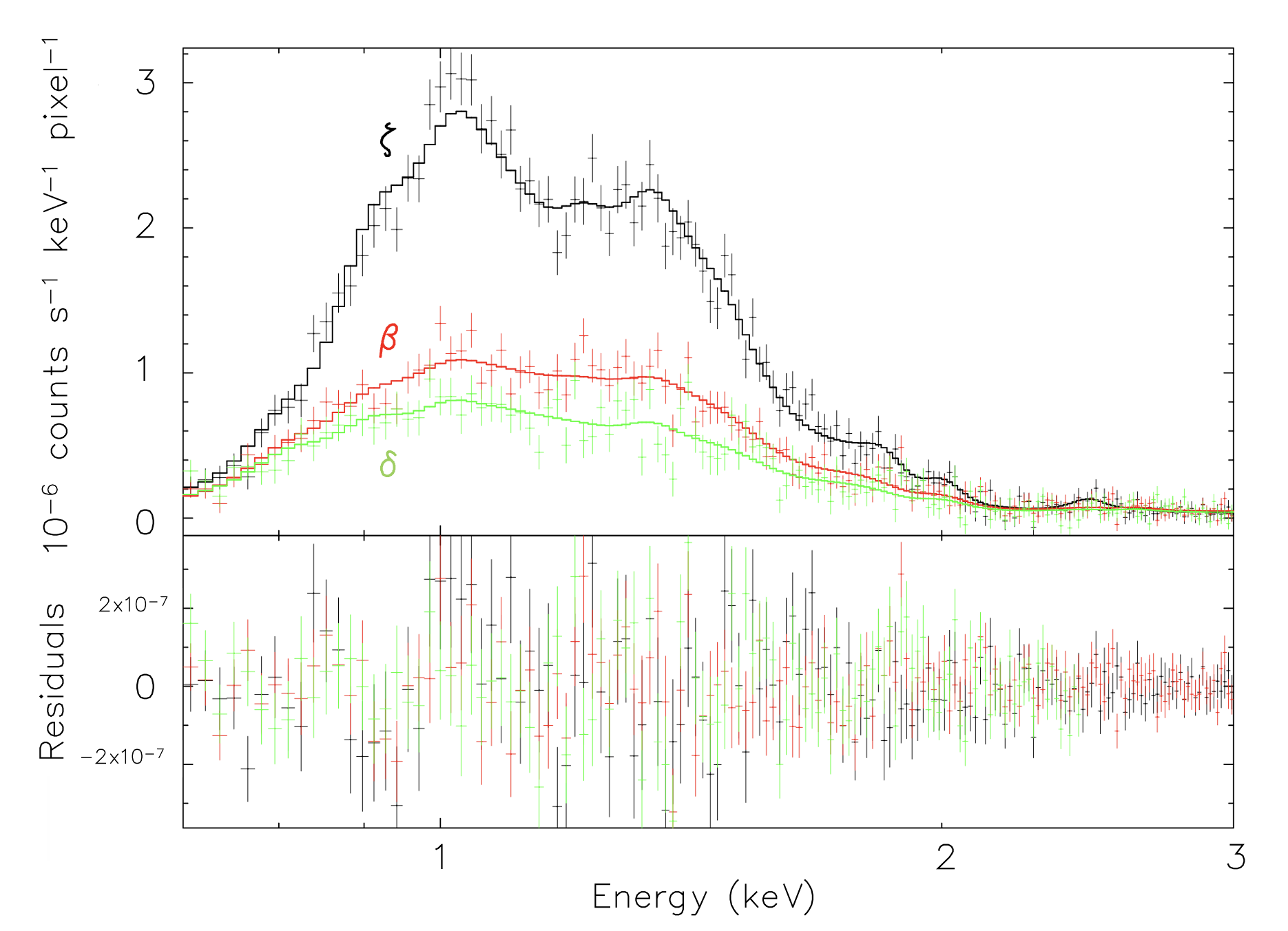}
    \caption{Spectral data from relatively bright ridges $\zeta$ (top), $\beta$ (middle), and $\delta$ (bottom) rescaled to account for differences in area.}
    \label{fig:spec_data}
\end{figure*}

The spectra from the five regions were fit simultaneously to determine the most likely physical model and set of parameters for each region (Table \ref{tab:simfit}).
When fitting the data, we find that there is little variation in the recombination timescale $\tau$ and the initial recombining plasma temperature across the seven regions, and we therefore link the regions together for these parameters.
We also fix the metal abundance to 1.0 in Solar units.
We evaluate the goodness-of-fit using the reduced chi-squared value (i.e., $\chi^2 / \mathrm{degrees \ of \ freedom}$), where a lower value of $\chi^2 / \mathrm{d.o.f.}$ indicates a better fit.
For each region, the model with the lowest $\chi^2 / \mathrm{d.o.f.}$ consists of a combination of RP and APEC components, each convolved with a Tuebingen-Boulder ISM grain absorption model (`TBabs'; \citealt{Wilms+00}), i.e. \texttt{tbabs}*(\texttt{rnei+apec}).
This model yields a $\chi^2 / \mathrm{d.o.f.}$ value of 1160/1062 = 1.092, which is significantly better than the other attempted models, such as a double APEC model (1921/1064 = 1.805). A comparison of the spectra for features $\beta$, $\delta$, and $\zeta$ fit using the RP+APEC model is shown in Figure \ref{fig:spec_data}.

\begin{deluxetable*}{lCCCCCCC}
\tablenum{2}
\tablecaption{Simultaneous Fit
\label{tab:simfit}}
\tablewidth{0pt}
\tablehead{
\colhead{Parameter} & \colhead{$\alpha$} & \colhead{$\beta$} & \colhead{$\gamma$} & \colhead{$\delta$} & \colhead{$\epsilon$} & \colhead{$\zeta$} & \colhead{$\eta$}
}
\startdata
\textbf{n$_H$} ($10^{22} \ \mathrm{cm}^{-2}$) & 0.79 \pm 0.04 & 0.86 \pm 0.04 & 0.92 \pm 0.07 & 0.75 \pm 0.07 & 0.95 \pm 0.04 & 0.92 \pm 0.03 & 0.93 \pm 0.03 \\
\textbf{RP kT}$_{\mathrm{init}}$ (keV) & \multicolumn{7}{c}{$0.35 \pm 0.01$ $\mathrm{(linked)}$}  \\
\textbf{RP kT} (keV) & 0.183 \pm 0.009 & 0.18 \pm 0.01 & 0.20 \pm 0.02 & 0.19 \pm 0.01 & 0.24 \pm 0.02 & 0.148 \pm 0.009 & 0.15 \pm 0.01 \\
\textbf{RP} $\boldsymbol{\tau}$ ($10^{10}$ s $\mathrm{cm}^{-3}$) & \multicolumn{7}{c}{$9 \pm 3$ $\mathrm{(linked)}$}\\
\textbf{APEC kT} (keV) & 2.4 \pm 0.4 & 2.8 \pm 0.8 & 1.8 \pm 0.4 & 2.8 \pm 0.8 & 1.05 \pm 0.06 & 1.22 \pm 0.09 & 1.49 \pm 0.08 \\
\textbf{APEC EM} ($10^{52} \ \mathrm{cm}^{-3}$) & 0.31 \pm 0.02 & 0.36 \pm 0.03 & 0.27 \pm 0.04 & 0.31 \pm 0.03 & 0.42 \pm 0.05 & 0.60 \pm 0.06 & 0.94 \pm 0.05 \\
\textbf{RP EM} ($10^{52} \ \mathrm{cm}^{-3}$) & 8.9 \pm 0.8 & 12 \pm 1 & 7 \pm 1 & 6 \pm 1 & 3.9 \pm 0.9 & 35 \pm 3 & 25 \pm 2 \\
\enddata
\end{deluxetable*}

\section{Interpretations and Discussion}
\label{sec:disc}

\subsection {Hypothesis: a Cylindrical Tunnel}

The IBDF feature contains two prominent bright ridges, regions $\beta$ and $\zeta$, which border a column of relatively low X-ray surface brightness (regions $\gamma$, $\delta$, and $\epsilon$). The linear morphology of these quasi-parallel ridges, coupled with the fact that ridge $\beta$ has a sharp western edge, but falls off less abruptly toward the east into region $\alpha$, leads us to hypothesize that the IBDF feature is a cylindrical, edge-brightened tunnel lying along the central axis of the southern chimney.  In keeping with the notion that the chimneys represent collimated plasma outflows from the Galactic center \citep{Ponti+19}, we regard this tunnel as the channel, or vent, along which the outflowing plasma is moving.
While no measurement yet exists of the outflowing plasma velocity within the putative tunnel, we argue that it must be at least as large as the velocity of the wind emanating from the overall Galactic center region, $\gtrsim 1000$ km s$^{-1}$ \citep{Carretti+13, Fox+15, Fujita23}.  
We assume that the outflow velocity of the plasma is highest along the central axis of the tunnel because the flow along the edges would be somewhat slowed by  turbulent interactions with the adjacent ambient medium. The relatively low X-ray brightness of the interior of the tunnel can then be attributed to a lower plasma density there via the continuity equation, coupled with the density-squared dependence of the X-ray emissivity of the plasma.

The relatively high emissivity of the tunnel walls, manifested as the ridges $\beta$ and $\zeta$, can be attributed to shocks that occur where the outflowing plasma impacts and compresses the surrounding ambient gas.  The particularly high brightness of ridge $\zeta$ is readily attributable to its orientation; it is slightly askew relative to the axis of the tunnel, so the outflowing plasma strikes this portion of the tunnel wall at a more direct angle, and therefore with more force, than it does elsewhere where the velocity vector of the outflowing plasma is presumably almost parallel to the walls of the tunnel.

We note that the detailed morphology of the tunnel appears to be somewhat more complex than a perfectly uniform cylinder, given the presence of a fainter ridge, $\delta$, projected near the center of the tunnel.  

\subsection {Hardness Ratio and Plasma Temperature Distribution}

The brightest regions in the IBDF coincide with the smallest hardness ratios when comparing 1.2-2.0 keV and 0.5-1.2 keV emission.
If the IBDF feature is indeed attributable to shocks propagating into the ambient medium, then this shock compression of the ``tunnel walls" (i.e. $\beta$ and $\zeta$) would naturally lead to higher densities.
This high density would in turn give rise to greater emissivity and therefore more rapid cooling, ultimately producing the bright but soft X-ray emission that we observe.
A plausible candidate for energy injection into this region is the series of ``hundred-year events" that have been inferred from moving X-ray echoes observed primarily in the fluorescent 6.4 keV iron line \citep{Ponti+10, Clavel+12, Ponti+13, Churazov+17, Chuard+18, Marin+23}.
If the cooling time of the IBDF plasma is greater than a few hundred years, then this sequence of hundred-year events, presuming that they occur with some regularity on few-hundred-year timescales, may effectively act as a continuous flow that continually replenishes the region.

Using the fit parameters for density and temperature from Table \ref{tab:simfit}, we can estimate the cooling timescale of the shocked plasma for comparison to this few-hundred-year timescale.
The characteristic cooling timescale of the plasma can be calculated using the thermal energy and emitted power of the plasma via the relation
\begin{equation}
    \tau_{\mathrm{cool}} = \frac{\frac{3}{2} \left(1 + \frac{n_{\mathrm{i}}}{n_{\mathrm{e}}}\right)kT}{\Lambda(t)n_{\mathrm{H}}}.
\end{equation}
Here $\Lambda(t)$ is the cooling function, which in this case essentially amounts to bolometric power normalized by emission measure. 
Using cooling curves provided by \cite{Sutherland+93}, we estimate $\Lambda(t) \sim 5 \times 10^{-30} \ \mathrm{W \ cm}^{3}$ based on temperature and approximate metal abundance.
Then assuming $n_{\mathrm{e}} = n_{\mathrm{i}} = n_{\mathrm{H}}$, we can estimate the cooling timescale for various possible geometries.
To get a lower limit, we first assume that emission is confined to a volume whose depth is equal to its angular width so that the dimensions of region $\beta$ are $0.1^\circ \times 0.01^\circ \times 0.01^\circ$, giving us an estimated cooling timescale in this region of $\gtrsim 200$ yr.
This estimate assumes a filamentary feature in the simplest possible case.
If, however, we are looking at a cylindrical tunnel with a diameter of about $0.1^\circ$, then the depth of region $\beta$ is represented by a chord on that cylinder lying approximately $0.025^\circ$ from the center. Thus, the depth is $\sim0.087^\circ$, and we calculate an estimated cooling timescale of $\gtrsim2000$ yr.

To maintain the observed structure, the non-equilibrium plasma of the IBDF must be continually replenished or heated on a timescale shorter than this cooling time, and the Sgr A$^*$ hundred-year events therefore provide a plausible mechanism for sustaining the heating and outflow of the IBDF plasma and counteracting this relatively short cooling timescale.

\subsection{Recombining Plasma}
\cite{Nakashima+13} report \textit{Suzaku} results indicating the presence of a recombining plasma south of the Galactic plane in a larger region that contains the IBDF. 
Our spectral analysis of the IBDF is indeed consistent with this result as a model containing a recombining plasma and a thermal APEC component produced acceptable fits with the lowest $\chi^2$ values.
We find that no single-plasma model alone is able to sufficiently fit the data, which leads us to introduce the additional APEC component in our RP+APEC and APEC+APEC models.
The improvement to the fit afforded by the additional APEC component to the RP model suggests that there may be an associated thermalized plasma mixed in with or adjacent to the bulk recombining plasma, or that the plasma may have a more-or-less continuous distribution of temperatures.
Comparing the temperatures in Table \ref{tab:simfit}, we see that the APEC component is at a higher temperature than the RP component for all 7 of the sub-regions in the fit. 
This may be an indication that the plasma responsible for the observed bright features is distinct from another plasma component sitting within or moving through the tunnel, or that the plasma of the IBDF has a highly variable temperature distribution.
Further spectral analysis of the IBDF will be required to provide a more accurate picture. Follow-up high-resolution spectroscopy with XRISM may be especially helpful in this effort.

The presence of a recombining plasma in the IBDF is also consistent with the Galactic outflow hypothesis. 
The morphology of the IBDF features, along with the trends in hardness ratio, suggest the existence of shocks in the region.
Strong shocks stemming from the outflow would propagate into the ambient interstellar medium at high velocity, compacting the gas and sustaining the recombining plasma environment.

\subsection{Relationship to Smaller- and Larger-Scale Features} 

A number of features observed on different scales merit consideration as possibly being related to the X-ray features in the IBDF.  As mentioned above, the X-ray chimneys, in which the IBDF is embedded, have been invoked as the channel through which plasma and cosmic rays generated in the vicinity of the Galactic center transit out to the Galactic halo, potentially provoking the $\gamma$- and X-ray emission arising in the large-scale Fermi and eRosita Bubbles \citep{Ponti+19,Ponti+21}.  

We also note that on the scale of a few parsecs, a linear X-ray feature close to the Galactic black hole has been identified and interpreted as a jet from the black hole \citep{Baganoff+03, Muno+08, LiZ+13, ZhuZhenlin+19}.  That putative jet is oriented in the same direction as the southern chimney and the linear features in the IBDF, but it has not been detected at distances exceeding a few parsecs from the black hole, apparently because its nonthermally-emitting particles have lost sufficient energy beyond that point to continue emitting a detectable flux of X-rays \citep{ZhuZhenlin+19}.  However, at very low radio frequencies, \citet{YZ+86a} \citep[see also][]{Kassim+86} reported the presence of a ridge of radio continuum emission extending from the Galactic center out to a distance of about 25 pc normal to the Galactic plane, that is, in the same direction as the hypothetical X-ray jet, but far short of the 220 pc distance of the IBDF from the Galactic black hole. Continued energy loss by the electrons in the hypothetical jet could lead to a fossil jet of low-energy particles responsible for the observed low-frequency synchrotron emission.  Eventually, if those particles are moving through the plasma in the chimney, they will thermalize with the plasma, and thus have no observable excess manifestation of their presence at the distance of the IBDF.  

While the placement and morphology of all the features mentioned here are very suggestive, further investigation is clearly needed to establish and elucidate the physical links between them.

\section{Summary}
\label{sec:summary}
The 1 Ms Chandra field centered at $l = 0.08^{\circ}, ~b = -1.42^{\circ}$ contains a bright X-ray structure having a striated linear morphology and hardness ratio trends suggestive of a cylindrical tunnel. 
The linear strands of X-ray emission are approximately perpendicular to the Galactic plane and sit neatly within the diffuse emission of the southern Galactic Center chimney, which is itself situated inside a wider shell of radio emission. 
Because the roughly cylindrical chimney is centered on the Galactic black hole, the natural hypothesis for the overall structure is that an X-ray emitting plasma generated at or in a broad region within $\sim$100 pc of the black hole is flowing out of central region through the chimney, and we hypothesize that the apparent tunnel that we report here is the central conduit for that plasma.
While our spectral analysis of the linear X-ray features does not conclusively constrain the metallicity or show an obvious trend in temperature, it does point toward the presence of a two-temperature plasma.
Such a plasma consisting of a thermal and a recombining component could be sustained from shocks due to the continuous outflow proposed in the chimney hypothesis.
The inferred tunnel-like morphology is best supported by the X-ray hardness ratios, which demonstrate a clear difference between the brighter ``walls" of the tunnel and the dimmer region in-between.
The bright outer regions coincide with the lowest hardness ratios, a possible indicator of higher density and more rapid cooling as a result of shock compression from outflowing plasma.

The presumably outflowing plasma in the tunnel is a strong candidate for the source of the Galactic wind, and perhaps for providing the particles and energy that are responsible for the gamma-rays emanating from the Fermi Bubbles and the X-rays arising in the eRosita Bubbles.  The biggest remaining open question is whether those galaxy-scale structures were created predominantly in a past major black-hole accretion event or whether they are the result of a sequence of frequent episodic energy releases within the Galaxy's central region. In the latter case, the observed plasma structures -- chimneys and tunnel -- are enduring features that carry the intermittent spurts of energy out to the large-scale bubbles.

One very useful future investigation would be to determine the plasma velocity (and gradient) within the chimneys, either by carrying out long-term X-ray proper motion studies or by measuring the Doppler shifts of X-ray lines with a high-resolution X-ray spectrometer.  The challenge with the former measurement is the limited resolution of X-ray imagers (with Chandra resolution, it would take about a decade to make a proper motion measurement of an unresolved feature moving at 1000 km s$^{-1}$ at the Galactic center, and the features reported here are partially resolved), while the challenge with the latter measurement is that the predominant direction of the plasma flow is likely to be perpendicular to our line of sight.  

\bigskip

{\bf Acknowledgements:} The work carried out for this project by UCLA participants was supported by NASA/SAO grant GO1-22138X. 
GP acknowledges financial support from the European Research Council (ERC) under the European Union’s Horizon 2020 research and innovation program “HotMilk” (grant agreement No. 865637) and support from Bando per il Finanziamento della Ricerca Fondamentale 2022 dell’Istituto Nazionale di Astrofisica (INAF): GO Large program.

\bibliography{tunnel}{}
\bibliographystyle{aasjournal}


\appendix

\section{Background}
\label{appendix:a}
In order to study diffuse X-ray emission, the non-X-ray background from the Chandra instruments must be removed from the data.
This background is primarily caused by cosmic ray impacts on the ACIS CCDs and nearby components, and therefore is not affected by the optical elements of the telescope \citep{Hickox+06}. 
To measure this background, several Chandra observations have been made with ACIS in the stowed position to block out external X-rays and ensure that any measured flux originates from within the telescope. 
We downloaded these datasets from the Chandra website\footnote{\url{https://cxc.harvard.edu/contrib/maxim/acisbg/}} and reprocessed them according to the standard procedure used for the science data. 

Because the effective area of ACIS for energies of 9-12 keV is nearly negligible, we begin with the assumption that essentially all of the flux in this band consists only of instrumental background.
We therefore first proceeded by assuming that the count rate in this band should be the same in the background data as it is in the science data.
Re-scaling the background count rate to account for exposure time, we would then subtract the background data from the science data.
However, in doing this, we found that the count rate becomes negative above 4 keV, which is indicative of an apparent over-subtraction.
Thus, we implemented a corrective re-scaling procedure.
We began by reducing the amplitude of the background data by a factor of 1.37. This factor was determined by assuming that prominent 7.5 keV Ni K$\alpha$ and 9.7 keV Au L$\alpha$ lines in the spectra were entirely due to internal instrumental emission, and then scaling the background such that these lines vanish.
Then, to ensure we had the optimal re-scaling factor, we chose a small uniform region on the edge of the IBDF to fit with a simple thermal model and then re-fit the region as we varied the background re-scaling factor.
This indicated that a factor of 1.31 from the original instrumental background data was the scaling that provided the best fit, as determined using the same criterion as in Section \ref{sec:spectra}.
We subsequently adopted this re-scaling factor throughout the rest of our analysis.

\end{document}